\documentclass[aps,prx,twocolumn, nofootinbib, epsfig,showpacs,floatfix, bibliography]{revtex4-2}
\usepackage{amsmath}
\usepackage{xcolor}
\usepackage{url}
\usepackage{physics}
\usepackage{graphicx}
\usepackage{amsfonts}
\usepackage{mathrsfs}
\usepackage{amsmath}
\usepackage{amssymb}
\usepackage{float}
\usepackage{bbold}
\usepackage[T1]{fontenc}
\usepackage[ttdefault=true]{AnonymousPro}
\usepackage[section]{placeins}
\usepackage[utf8x]{inputenc}
\usepackage{times}
\usepackage[colorlinks=true, linkcolor=blue, urlcolor=blue, citecolor=blue]{hyperref}
\makeatletter
\newcommand*{\rom}[1]{\expandafter\@slowromancap\romannumeral #1@}
\makeatother
\usepackage{multirow}
\hypersetup{
  colorlinks=true,
  citecolor=blue,
  linkcolor=blue,
  urlcolor=blue}

\begin{document} 

\title{Measurement-induced entanglement transition in chaotic quantum Ising chain}
 
\author{Manali Malakar\textsuperscript{1}, Marlon Brenes\textsuperscript{2}, Dvira Segal\textsuperscript{2,3}, and Alessandro Silva\textsuperscript{1}}

\affiliation{\textsuperscript{1}International School for Advanced Studies (SISSA), Via Bonomea 265, 34136 Trieste, Italy\\
\textsuperscript{2}Department of Physics and Centre for Quantum Information and Quantum Control, University of Toronto, 60 Saint George St., Toronto, Ontario, M5S 1A7, Canada \\
\textsuperscript{3}Department of Chemistry, University of Toronto, 80 Saint George St., Toronto, Ontario, M5S 3H6, Canada}

\date{\today}

\begin{abstract}
We numerically investigate the robustness against various perturbations of measurement-induced phase transition in monitored quantum Ising models in the no-click limit, where the dynamics is described by a non-Hermitian Hamiltonian. We study perturbations that break the integrability and/or the symmetry of the model, as well as modifications in the measurement protocol, characterizing the resulting chaos and lack of integrability through the  Dissipative Spectral Form Factor (DSFF). We show that while the measurement-induced phase transition and its properties appear to be broadly insensitive to lack of integrability and breaking of the $\mathbb{Z_2}$ symmetry, a modification of the measurement basis from the transverse to the longitudinal direction makes the phase transition disappear altogether. 
\end{abstract}


\maketitle

\section{Introduction}

One of the most important characteristics of open quantum systems is 
their inherent stochasticity, either associated to dephasing and dissipation, or to the measurement process~\cite{milburn}. Such stochasticity makes it hard to repeat the same time evolution twice, meaning that the comparison between theory and experiments can be done only upon averaging over many \it quantum trajectories\rm. When the averages are those of local operators or correlation functions, which share the property of being linear in the corresponding density matrix, the physics is effectively described by standard Gorini-Kossakowski-Sudarshan-Lindblad (GKSL) equation \cite{manzano}. This is not the case for physical quantities that zoom  in the structure of individual wave functions, such as the entanglement entropy~\cite{Ent_1,Ent_2,Ent_3,Ent_4,calabrese,kitaev_1} or the quantum Fisher information~\cite{metrology}: in this case it was shown that increasing the measurement rate may lead to phase transitions in the entanglement properties of the system which are not detected by the GKSL equation. These phenomena have been termed \it measurement-induced phase transitions \rm (MIPT) \cite{fisher_1, fisher_2, nahum, chan, ippoliti, huse,sierant,altman,zhu,romito,daley,turkeshi_1,turkeshi_2,turkeshi_3,ashida, diehl_1,diehl_2,diehl_3,turkeshi_4,turkeshi_5,turkeshi_6,silva_1,silva_2,kalsi,kells,tiwari,zhang}. 

Since MIPT refer to the entanglement structure of individual wave functions, one might hope that focusing on the physics of a single, simple yet typical trajectory would give information about the overall transition. This is the spirit that leads to the study, in the context of weakly measured many-body systems, of the so called \it no-click limit\rm~\cite{turkeshi_5,turkeshi_9}: a peculiar trajectory where the measurement fails along the entire stochastic evolution. The \it no-click limit \rm is particularly simple because the wave function dynamics is described by the time evolution with respect to a many-body non-Hermitian Hamiltonian~\cite{ueda0, ruhman_fm,schiro,gullans,chen, zhang_carrasquilla_kim}, typically the sum of a Hermitian counterpart plus an imaginary term associated with the measurement. This simplification, which makes problems exactly solvable in some instances, comes at a price: intuition developed for standard many-body systems cannot be automatically translated into the non-Hermitian realm since mathematically the spectra and eigenstates of the latter do not posses the same stability against perturbations of their Hermitian counterparts~\cite{ueda0}. Nevertheless, some interesting parallels have been observed: for various integrable many-body models in one dimension it was observed that the scaling of the \it stationary state \rm entanglement entropy appears to be controlled by the gap in the \it imaginary part \rm of the spectrum, analogous to what happens for the ground state entanglement of the Hermitian counterpart and the gap in its \it real \rm spectrum~\cite{silva_2,turkeshi_2}. However, it is unclear a priori how broadly applicable these observations are, particularly when the model is made generic through perturbations that break integrability.

In this paper, we take a step towards addressing the above question by numerically studying the long-time steady state in the \it no-click limit \rm of the quantum Ising chain under continuous measurements of the local transverse (and longitudinal) magnetization, while implementing different strategies to disrupt its integrability. To assess how the scaling properties of entanglement respond to the breaking of integrability and symmetry,  we drive the system into the chaotic regime, initially by adding a strong ferromagnetic interaction between the next-nearest neighbor (NNN) spins, followed by the introduction of a uniform longitudinal field that further breaks the $\mathbb{Z}_{2}$ symmetry of the Hamiltonian under consideration. While the results obtained in the chaotic regime appear to be consistent with a persistent transition and a connection between the gap in the imaginary part of the spectrum and the scaling of entanglement entropy, it is found that monitoring the local spins along the longitudinal direction leads to a complete disruption of the gapped-to-gapless transition, as witnessed from both the entanglement scaling with respect to system size and the spectral properties of the system as compared to the previous scenario.

The rest of the paper is organized as follows. In Sec.~\ref{Model_and_Method}, we outline the properties of the non-Hermitian system arising from local measurements of the transverse field in the standard quantum Ising chain under the \it no-click limit. \rm Following this, we incorporate various integrability-breaking terms into the system and report our findings on entanglement scaling and spectral properties in Sec.~\ref{Transverse}. Subsequently, in  Sec.~\ref{Longitudinal}, we discuss the implications of monitoring the local spins along the longitudinal direction and finally we summarize and conclude our results in Sec.~\ref{Conclusion}.\\ 

\section{Model, Measurement Protocols and the no-click limit}
\label{Model_and_Method}
The starting point of our analysis is the one-dimensional quantum Ising model described by the following Hamiltonian:
\begin{eqnarray}
\hat{\mathcal{H}}_{\mathrm{QI}}=-J\sum\limits_{i=1}^{L}\hat{\sigma}^{z}_{i}\hat{\sigma}^{z}_{i+1}-h\sum\limits_{i=1}^{L}\hat{\sigma}^{x}_{i}
\label{QIC}
\end{eqnarray}
where $\hat{\sigma}^{\alpha}$ with $\alpha \in \{x, y, z\}$ stand for the Pauli spin matrices, $L$ is the number of lattice sites, $J$ represents the nearest neighbor interaction, and $h$ denotes the transverse field. In our study, we consider the system with periodic boundary conditions (PBC), and therefore, the Hamiltonian in Eq.~\eqref{QIC} exhibits translational symmetry, $\mathbb{Z}_2$ spin-inversion symmetry and reflection symmetry; the last of which involves spatial reflection about the center of the chain. We resolve each of these symmetries to reduce the dimension of the Hilbert space to a given subsector. Furthermore, we scale all energies with respect to the nearest neighbor interaction and set $J=1$ as the energy scale, unless otherwise specified. 

Below we will be interested in studying the physics of non-Hermitian versions of Eq.~(\ref{QIC}) which can be interpreted as effective Hamiltonian $\mathcal{\hat{H}}_{\rm eff}$ for the \it no-click limit \rm of a stochastic Schr\"{o}dinger equation (SSE). Let us first see how the latter emerges: a measurement apparatus weakly measuring the system can be characterized using Positive Operator Valued Measures (POVMs) \cite{milburn,wiseman,jacobs,svensson,ahnert,brun}, described by suitable Kraus operators $\hat{A}_{n}$  satisfying the condition $\sum_{n}\hat{A}^{\dagger}_{n}\hat{A}_{n}=\mathbb{1}$, which can further be written in the site-decoupled form  as: $\hat{A}_{n}=\bigotimes_{i=1}^{L} \hat{A}^{(r)}_{i}$, with $r \in \{0,1\}$. Let us start by considering the following Kraus operators \cite{milburn}
\begin{subequations}
\begin{align}
\hat{A}^{(0)}_{i} &= \hat{M}^{x}_{i-}+\sqrt{1-\gamma dt}\hat{M}^{x}_{i+}\\
\hat{A}^{(1)}_{i} &= \sqrt{\gamma dt}\hat{M}^{x}_{i+}
\label{Kraus}
\end{align}
\end{subequations}
where $\hat{M}^{x}_{i\pm}=(\mathbb{1}\pm \hat{\sigma}^{x}_{i})/2$ represent the local projectors onto the eigenstates of $\hat{\sigma}^{x}$: $\hat{\sigma}^{x}|\pm\rangle=\pm|\pm\rangle$ and $\gamma$ denotes the measurement rate. The probabilities associated with the two possible measurement outcomes are $p_{1}=\gamma dt \langle \hat{M}^{x}_{i+} \rangle_{t}$  \big(where $\langle \hat{M}^{x}_{i+} \rangle_{t}=\langle \psi_{t}|\hat{M}^{x}_{i+}|\psi_{t}\rangle$\big) and $p_{0}=1-p_{1}$. Undergoing occasional yet instantaneous measurements on the quantum states, the dynamics of the quantum trajectories are now regulated by a Stochastic Schr\"{o}dinger equation \cite{molmer,daley_sse,plenio,milburn}, given by
\begin{eqnarray}
d|\psi_{t}\rangle&=&-\imath \hat{\mathcal{H}}_{\rm QI}dt|\psi_{t}\rangle  - \frac{\gamma}{2}\sum \limits_{i}\bigg(\hat{M}^{x}_{i+}-\langle \hat{M}^{x}_{i+} \rangle_{t} \bigg) dt|\psi_{t}\rangle \notag\\
&+&\sum \limits_{i}\delta N^{i}_{t} \Bigg(\frac{\hat{M}^{x}_{i+}}{\sqrt{\langle \hat{M}^{x}_{i+} \rangle_{t}}}-1 \Bigg)|\psi_{t}\rangle
\label{SSE}
\end{eqnarray}
where $\delta N^{i}_{t}\in \{0,1\}$ are the local Poisson processes, satisfying $\overline{\delta N^{i}_{t}}=\gamma dt\langle \hat{M}^{x}_{i+} \rangle_{t}$. Whenever $\delta N^{i}_{t}=1$, the measurement apparatus clicks, causing the quantum state to undergo a discontinuous jump along $|+\rangle$ \cite{turkeshi_5}. In the specific case $\delta N^{i}_{t}=0$ at each site, the time evolution of the system is written as
\begin{eqnarray}
|\psi_{t}\rangle = \frac{e^{-\imath \hat{\mathcal{H}}_{\rm eff}t}|\psi_{0}\rangle}{||e^{-\imath \hat{\mathcal{H}}_{\rm eff}t}|\psi_{0}\rangle||}
\label{Time_Evolution}
\end{eqnarray}
and it is governed by a non-Hermitian Hamiltonian
\begin{eqnarray}
\hat{\mathcal{H}}_{\rm eff}= -\sum\limits_{i=1}^{L}\hat{\sigma}^{z}_{i}\hat{\sigma}^{z}_{i+1}-\big(h+ \imath \frac{\gamma}{4}\big)\sum\limits_{i=1}^{L}\hat{\sigma}^{x}_{i}.
\label{H_eff}
\end{eqnarray} 
This is the so called \it no-click limit\rm. Since $\hat{\mathcal{H}}_{\rm eff}$ is non-Hermitian, the eigenvalues are in general complex. The normalization in Eq.~(\ref{Time_Evolution}) in turn implies that, writing the initial condition $|\psi_{t}\rangle=\sum_j\;c_j |\eta_j\rangle$ where $|\eta_j\rangle$ are the right eigenvectors of $\hat{\mathcal{H}}_{\rm eff}$ with eigenvalues $\Lambda_j$, only the states with the largest ${\rm Im}[\Lambda_j]$ survive in the long-time limit.

%%%%%%%%%%%%%%%%Figure 1%%%%%%%%%%%%%%%%%%%%%%
\begin{figure}[H]
\centering
\includegraphics[width=0.9\columnwidth]{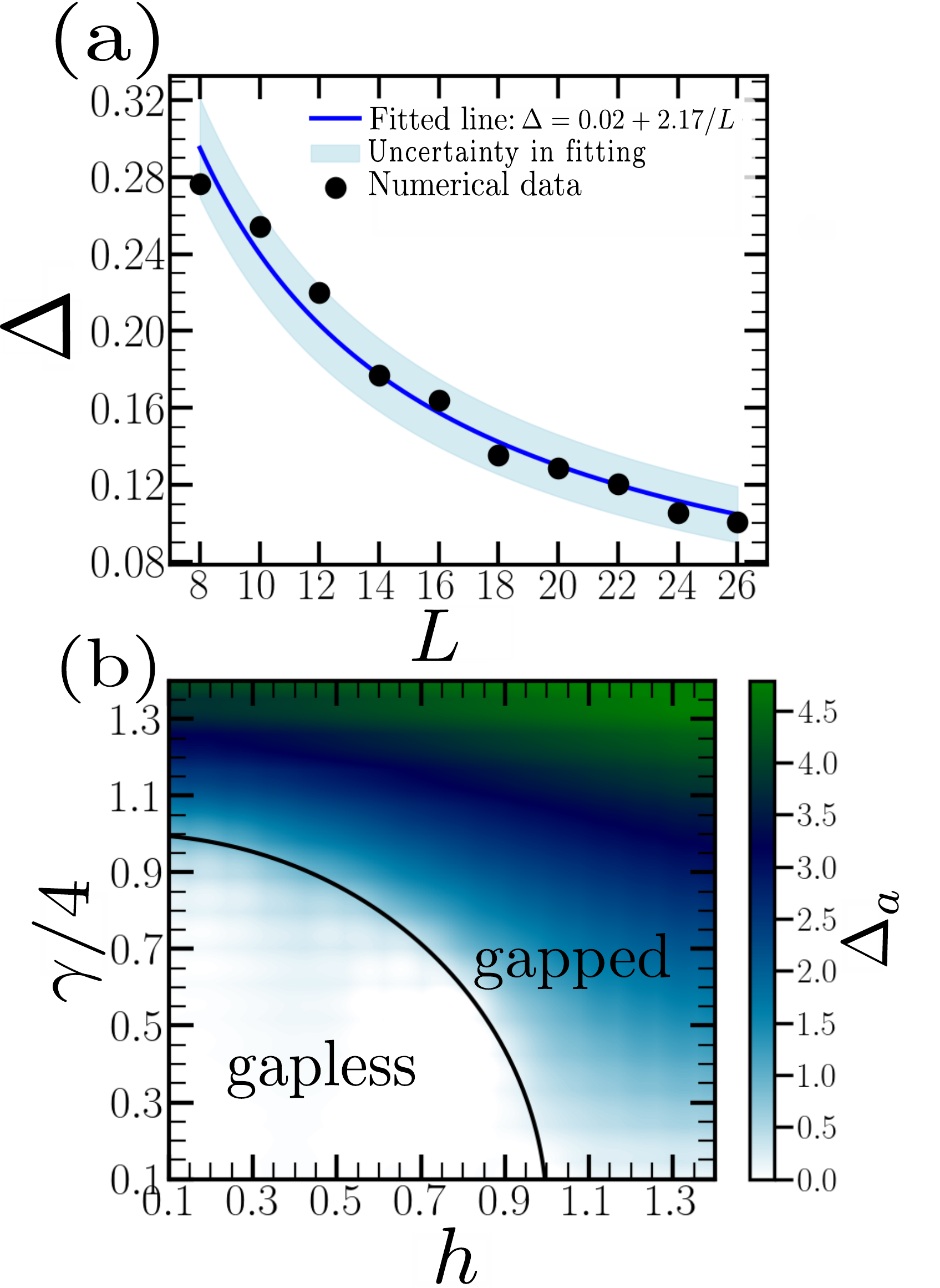}
\caption{(a) The variation of the spectral gap $\Delta$ with respect to the system size $L$ is depicted for $h=0.3$ and $\gamma=0.8<\gamma_{c}(h)\sim 3.82$. We perform a hyperbolic fitting (blue solid line) on our numerical data (black circles), gradually increasing the system size up to $L=26$, to extract the asymptotic gap $\Delta_{a}$ at the thermodynamic limit. The blue shaded region represents the uncertainty ($\sigma_{a}$ and $\sigma_{b}$) in the fitted parameters $a$ and $b$ for the hyperbolic fit  $f(L)=a+b/L$. (b) The phase diagram in the $\gamma$-$h$ plane, obtained numerically by extracting $\Delta_{a}$, illustrates a smooth transition from the gapless to gapped phase of the decay mode $\Gamma$. The critical boundary $\gamma_{c}(h)=4\sqrt{1-h^{2}}$, derived analytically, is depicted by the black solid line.}
\label{Fig1}
\end{figure}
%%%%%%%%%%%%%%%%%%%%%%%%%%%%%%%%%%%%%%%%%%%%%%

The effective Hamiltonian in Eq.~(\ref{H_eff}) can be diagonalized  by a Jordan-Wigner transformation 
in terms of non-Hermitian quasiparticles featuring complex spectra: $\Lambda_{k}=E_{k}+\imath \Gamma_{k}$
with a finite imaginary part $\Gamma_{k}$ \cite{turkeshi_4}. In the ferromagnetic phase ($h<h_c=1$) the imaginary part of the spectrum $\Lambda_k$ has a transition at $\gamma_{c}(h)=4\sqrt{1-h^2}$ from gapless phase to a gapped one at momentum $k^{*}=\arccos{h}$. The critical measurement rate $\gamma_c$ at which this transition occurs is directly linked to a change in its entanglement properties as well: for $\gamma<\gamma_c$ the scaling of the entanglement entropy with subsystem size is logarithmic while for $\gamma>\gamma_c$ it saturates to a constant~\cite{turkeshi_4,turkeshi_9}.

Since in the following we will be considering the physics of perturbed non-Hermitian Hamiltonians using exact diagonalization, it is first of all interesting to ascertain whether this method can detect spectral and entanglement transitions as the ones described above. To achieve this, we numerically diagonalize Eq.~\eqref{H_eff} and compute the entanglement entropy of the steady state as $t \to \infty$. This steady state is identified as the eigenstate corresponding to the largest imaginary part of the spectrum. For the \textit{integrable} Hamiltonian in Eq.~(\ref{H_eff}), this state is referred to as the non-Hermitian vacuum, denoted by $|\emptyset_{\eta} \rangle$, where $\eta$ represents the non-Hermitian quasiparticles. 
The second task is to compute the spectral gap $\Delta$. In this context, it is important to mention that the present analysis is confined to the zero-momentum sector of the Hamiltonian, where the quasi-particles emerge only in pairs at momentum $\pm k$. Since for finite-size systems, locating the specific momentum where the gap closes becomes challenging due to momentum discretization, we observe significant fluctuations in the gaps between eigenvalues with system size $L$.  
To smooth out finite-size fluctuations we therefore determine the difference between the largest imaginary part of the spectra, and the average of the second and third largest ones (first and second excited states, respectively): $\Delta=\Gamma^{0l}-(\Gamma^{1l}+\Gamma^{2l})/2$.

As illustrated in Fig.~\ref{Fig1}(a), the gap $\Delta$ varies with system size following a $1/L$ law, and is expected to saturate to a value $\Delta_a$ in the thermodynamic limit. To determine $\Delta_a$ in this limit, we employ hyperbolic scaling to fit our data and subsequently extract the asymptotic gap through hyperbolic regression. As evident from Fig.~\ref{Fig1}(a), at $h=0.3$ and $\gamma(=0.8)<\gamma_{c}(h)(\sim 3.82)$, the asymptotic gap $\Delta_{a}$ approaches zero ($\sim 0.02$) in the thermodynamic limit, consistent with the analytical predictions. Moreover, as depicted in Fig.~\ref{Fig1}(b), we numerically construct a phase diagram based on the calculation of $\Delta_{a}$ in the $\gamma$-$h$ plane, exhibiting a smooth crossover from the gapless to the gapped phase, marked by the transition boundary (black solid semicircle), defined as $\gamma_{c}(h)=4\sqrt{1-h^{2}}$.
%%%%%%%%%%%%%%%%Figure 2%%%%%%%%%%%%%%%%%%%%%%
\begin{figure}[ht]
\centering
\includegraphics[width=0.85\columnwidth]{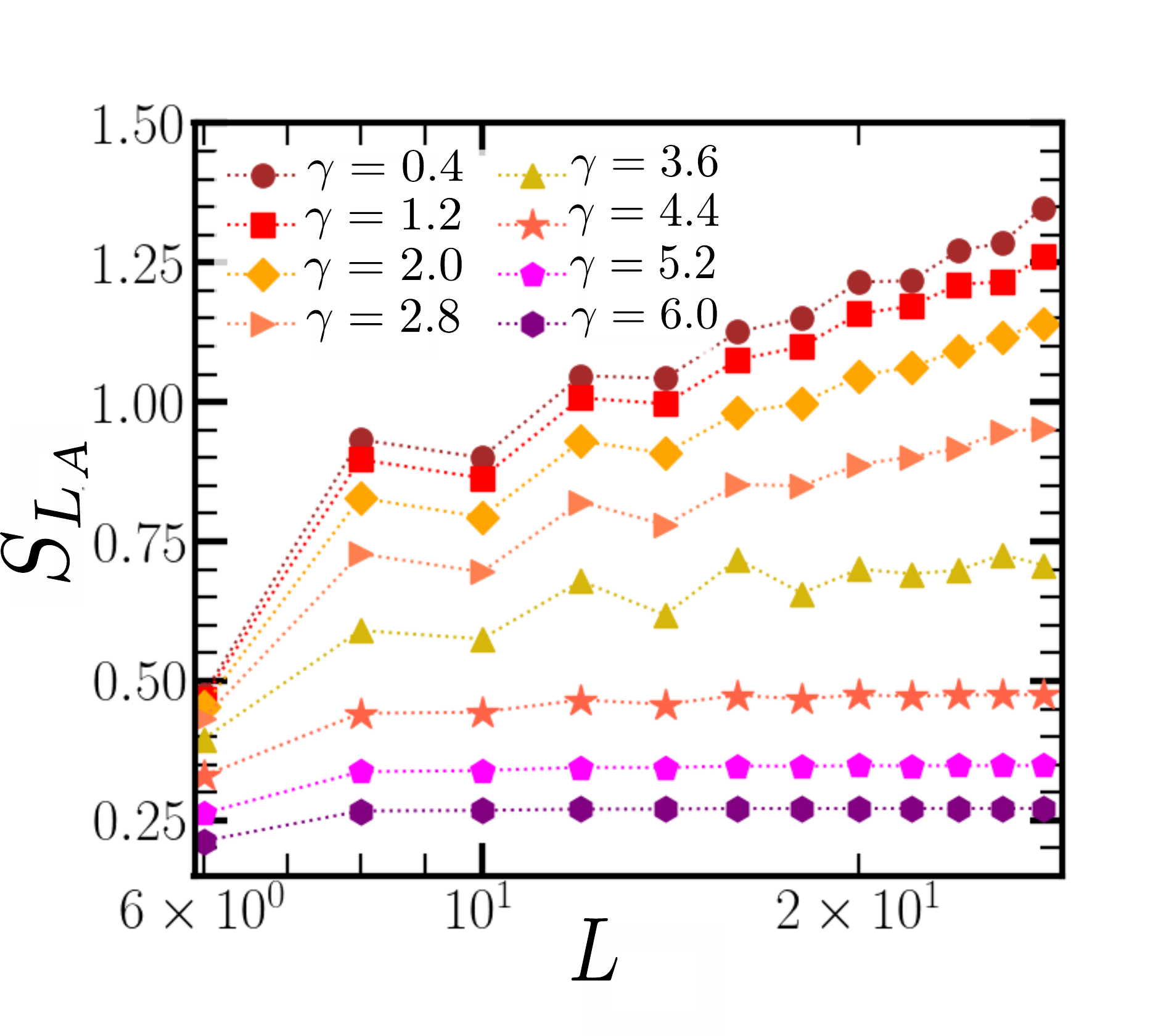}
\caption{Scaling of entanglement $S_{L_{\rm A}}$ of subsystem $L_{\rm A}=L/4$ as a function of $L$ is shown for transverse field, $h=0.2$ with increasing $\gamma$. For $\gamma<\gamma_{c}(h) (\sim 3.92)$, $S_{L_{A}}$ scales logarithmically with $L$, whereas for $\gamma>\gamma_{c}(h)$, it becomes constant, indicative of an area law. Here, we consider system sizes up to $L=28$.}
\label{Fig2}
\end{figure}
%%%%%%%%%%%%%%%%%%%%%%%%%%%%%%%%%%%%%%%%%%%%%%

Next, we examine the transition in entanglement scaling as the rate of measurement increases. We calculate the bipartite entanglement entropy (EE) of the stationary state at $t\to \infty$, for which the entanglement properties are fully encoded in the non-Hermitian vacuum $| \emptyset_{\eta} \rangle$ \cite{silva_2}. To evaluate the entanglement of the vacuum state, we compute the reduced density matrix $\hat{\rho}_{A}$ of the subsystem $L_{A}=L/4$ through the partial trace operation $\hat{\rho}_{A}={\rm Tr}_{B}\big[\hat{\rho}\big]$. Subsequently, the EE can be numerically evaluated from the von Neumann entropy
\begin{eqnarray}
S_{L_{A}}=-\sum\limits_{i}\lambda^{A}_{i}{\rm ln} (\lambda^{A}_{i}),
\label{Entanglement}
\end{eqnarray}
where $\lambda^{A}_{i}$ are the eigenvalues of the reduced density matrix $\hat{\rho}_{A}$.
In Fig.~\ref{Fig2}, we demonstrate how the entanglement entropy $S_{L_{\rm A}}$ of the vacuum state scales with system size $L$ as the measurement rate $\gamma$ increases. Within the  region of gapless spectrum for $\gamma<\gamma_{c}(h)$, entanglement grows logarithmically with $L$ along with some finite-size fluctuations. For $\gamma>\gamma_c$ it becomes constant, suggesting an area law behavior, associated with the gap opening in the imaginary part of the spectrum~\cite{silva_2}.\\

In the next section, we investigate the spectral and entanglement properties of the non-Hermitian Hamiltonian under the influence of integrability as well as symmetry-breaking terms.
%under which the evolved state at time $dt$ can be written as:
%\begin{eqnarray}
%|\psi(t+dt)\rangle=\frac{\hat{A}_{n}|\psi(t)\rangle}{\sqrt{\langle\psi(t)|\hat{A}^{\dagger}_{n}\hat{A}_{n}|\psi(t)\rangle}}
%\label{state_evolution}
%\end{eqnarray}
  
\section{Breaking of Integrability}
\label{Transverse}

Let us now turn to the study of the effects of perturbations that break the \it integrability \rm of the quantum Ising chain described by Eq.~\eqref{QIC} and Eq.~\eqref{H_eff}. We will focus on two tasks: the first is to investigate the  impact of perturbations on the overall spectrum and its deviation from the integrable limit by studying the Dissipative Spectral Form Factor (DSFF). The second task will be to study the stability upon perturbations of the entanglement and spectral transitions observed in the integrable model. In particular, we will study strong integrability-breaking terms, pushing the system into the \it chaotic \rm regime and look at the above-mentioned features.

\subsection{MODEL \rom{1}}
\label{sec:model1}

%%%%%%%%%%%%%%%%Figure 3%%%%%%%%%%%%%%%%%%%%%%
\begin{figure*}
\centering
\includegraphics[width=0.85\textwidth]{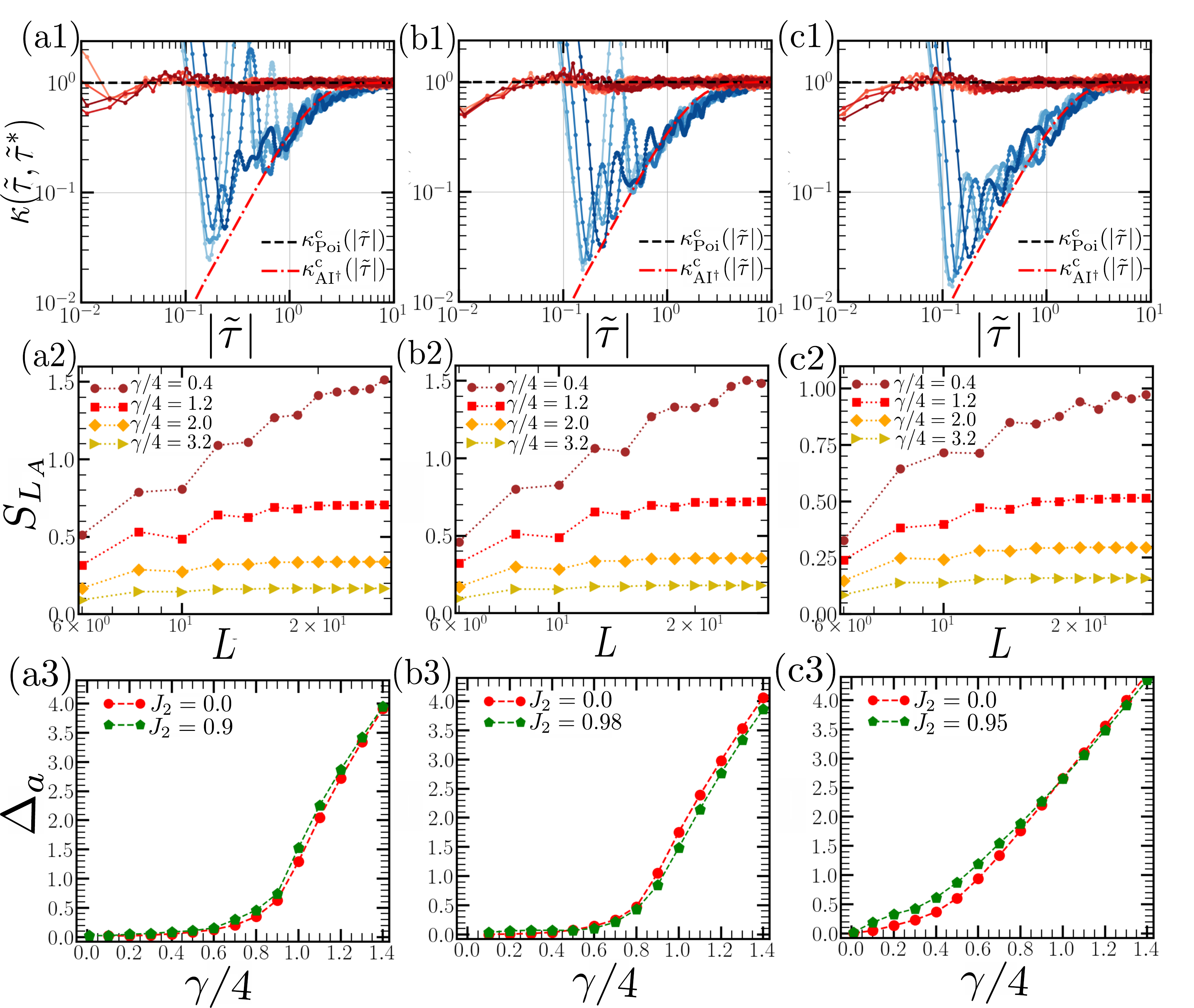}
\caption{\textit{First row:} (a1-c1) The DSFF $\kappa(\tilde{\tau},\tilde{\tau}^{*})$ is plotted as a function of $|\tilde{\tau}|$ for both the non-integrable chaotic limit (depicted in blue shades) and the near-integrable limit (represented in red shades) at specific $h$ values. The parameters chosen for the non-integrable chaotic limit are: (a1) $J_2 = 0.9$, $h = 0.3$, (b1) $J_2 = 0.98$, $h = 0.5$, and (c1) $J_2 = 0.95$, $h = 0.95$, and for the near-integrable limit, $J_2 = 0.01$. The different shades, varying from lightest to darkest, correspond to different $\theta$ values within the range $[\pi/18, 6\pi/18]$, increasing in steps of $\pi/9$. The black dashed line represents the DSFF for the uncorrelated Poissonian spectrum $\kappa^{c}_{\rm Poi}(|\tilde{\tau}|)=1$, while the red dash-dotted line, taken from Ref.~\cite{kulkarni_dsff}, depicts the DSFF $\kappa^{c}_{\rm AI^{\dagger}}(|\tilde{\tau}|)$ derived from random matrices belonging to the AI† symmetry class. The DSFF is calculated for fixed measurement rate $\gamma/4=0.4$ and the simulation is performed by averaging over 2000 random realizations of the system, each with $L = 14$. \textit{Second row:} (a2-c2) Entanglement entropy $S_{L_{A}}$ is plotted as a function of $L$ for (a2) $J_2 = 0.9$, $h = 0.3$, (b2) $J_2 = 0.98$, $h = 0.5$, and (c2) $J_2 = 0.95$, $h = 0.95$, with various strengths of $\gamma$ as indicated in the legend. \textit{Third row:} (a3-c3) The asymptotic gap $\Delta_{a}$ is plotted with increasing $\gamma$ at (a3) $h=0.3$, (b3) $h=0.5$, and (c3) $h=0.95$, both in the integrable ($J_{2}=0$, red circles) and chaotic limit (large $J_{2}$, green pentagons). We have taken  the system size up to $L=26$ to compute $\Delta_{a}$.}
\label{Fig3}
\end{figure*}
%%%%%%%%%%%%%%%%%%%%%%%%%%%%%%%%%%%%%%%%%%%%%%

In the first perturbed model we include a strong ferromagnetic next-nearest neighbor (NNN) interaction, resulting in the following Hamiltonian:
\begin{eqnarray}
\hat{\mathcal{H}}_{\rm \rom{1}}=-\sum\limits_{i=1}^{L}\hat{\sigma}^{z}_{i}\hat{\sigma}^{z}_{i+1}-J_{\rm 2}\sum\limits_{i=1}^{L}\hat{\sigma}^{z}_{i}\hat{\sigma}^{z}_{i+2}-h\sum\limits_{i=1}^{L}\hat{\sigma}^{x}_{i}
\label{QIC_NNN}
\end{eqnarray}
where $J_{2}$ denotes the NNN interaction. The Hamiltonian in Eq.~\eqref{QIC_NNN} with periodic boundary conditions retains the same symmetries outlined in Eq.~\eqref{QIC}. However, with the inclusion of $J_{2}\neq 0$, the system is no longer integrable.

We then couple the system with a measurement apparatus that measures the local spins $\hat{\sigma}^{x}_{i}$ along the direction of the transverse field, following the same protocol outlined in Sec.~\ref{Model_and_Method}. As a remnant of the monitoring process in the \it no-click limit, \rm the dynamics of the system is controlled by the resulting non-Hermitian Hamiltonian:
\begin{eqnarray}
\hat{\mathcal{H}}^{\rm \rom{1}}_{\rm eff}=&-&\sum\limits_{i=1}^{L}\hat{\sigma}^{z}_{i}\hat{\sigma}^{z}_{i+1}-J_{\rm 2}\sum\limits_{i=1}^{L}\hat{\sigma}^{z}_{i}\hat{\sigma}^{z}_{i+2}\notag \\
&-&\big(h+\frac{\imath \gamma}{4}\big)\sum\limits_{i=1}^{L}\hat{\sigma}^{x}_{i}
\label{H_eff_NNN}
\end{eqnarray}
where $\gamma$ represents the rate of measurement.

The Hamiltonian in Eq.~\eqref{H_eff_NNN} is neither integrable nor Hermitian. 
We therefore have to first verify whether the strength of the integrability-breaking term (in this case, $J_{2}$) is adequate to drive the system away from the integrable limit. In Hermitian systems this is done by studying the transition from Poisson to Wigner-Dyson statistics following the Bohigas-Giannoni-Schmit conjecture \cite{WD}. In non-Hermitian systems, the hallmark of chaos can be anticipated through the Dissipative Spectral Form Factor (DSFF) of non-Hermitian random matrices \cite{prosen_dsff,kulkarni_dsff}, as defined by
\begin{eqnarray}
\kappa(\tau,\tau^{*}) = \frac{1}{\mathcal{N}}\bigg \langle \bigg| \sum \limits^{\mathcal{N}}_{n=1} e^{\imath (z_{n}\tau^{*}+z^{*}_{n}\tau)/2} \bigg|^2 \bigg \rangle.
\label{DSFF}
\end{eqnarray}
Here, $z_{n}$ denotes the complex spectrum of the non-Hermitian Hamiltonian, and $\tau$ represents the complex time $\tau=|\tau|e^{\imath \theta}$. The summation is performed over the entire spectrum of the complex eigenvalues of an $\mathcal{N}$-dimensional matrix. The expression of $\kappa(\tau,\tau^{*})$ in Eq.~\eqref{DSFF} arises from the Fourier transformation of the two-point correlation function of the complex spectral density $\langle \rho(z_{1})\rho(z_{2}+\omega)\rangle$. In the quantum chaotic region, the DSFF, as a function of $|\tau|$, displays a universal dip-ramp-plateau behavior, indicating spectral correlations observed in non-Hermitian random matrix ensembles belonging to specific universality classes. Conversely, in the absence of chaos, such a ramp disappears, and the DSFF saturates to 1 after an initial dip, indicating Poissonian statistics of uncorrelated spectrum \cite{prosen_dsff,kulkarni_dsff}. In this context, it is crucial to determine the correct symmetry class of the non-Hermitian matrix under consideration. Our Hamiltonian exhibits a transposition symmetry, $\hat{H}=\hat{H}^{T}$, thereby belonging to the ${\rm AI^{\dagger}}$ non-Hermitian universality class~\cite{ueda}.

In Fig.~\ref{Fig3}(a1-c1), we illustrate the behavior of the DSFF, computed over the entire spectrum of complex eigenvalues for various combinations of $J_{2}$ and $h$, at specific $\theta$ and as a function of $|\tilde{\tau}|$. To facilitate the comparison between the numerically obtained DSFF and the behavior predicted by non-Hermitian Random Matrix Theory (RMT), we employ a rescaling of the time axis: $\tilde{\tau}=\tau/\tau_{H}$, where $\tau_{H}\sim \sqrt{\mathcal{N}}$ denotes the Heisenberg time (see Ref.~\cite{kulkarni_dsff}). Only a sufficiently large measurement rate $\gamma$ can induce adequate non-Hermiticity which leads, upon breaking integrability, to a favorable comparison to the DSFF in the symmetry class considered. As observed in Fig.~\ref{Fig3}(a1-c1), for larger values of the NNN interaction $J_{2}$ (the blue shaded lines), the DSFF aligns closely with that obtained from the non-Hermitian random matrix ensembles belonging to the ${\rm AI^{\dagger}}$ symmetry class, indicating the presence of chaos. Whereas, for $J_{2}\sim 0.01$ (the red shaded lines), the complex spectrum becomes uncorrelated, displaying a Poissonian behavior. 

In order to examine the implications of chaos on the steady-state entanglement we first isolate the eigenstate with largest ${\rm{Im}}[\Lambda_k]$ and plot the entanglement entropy $S_{L_A}$ of subsystem $L_{A}=L/4$ as the system size increases for different $\gamma$, particularly under the strong influence of $J_{2}$. In Fig.~\ref{Fig3}(a2-c2), it can be observed that with a relatively weak $\gamma$, the EE exhibits logarithmic growth as  $L$ increases, up to finite-size fluctuations. On the other hand, for larger $\gamma$, the entanglement becomes constant with system size, consistent with a transition from logarithmic to area law with increasing rate of measurement. It appears that the gap in the imaginary part of the spectrum has a behaviour roughly independent on the breaking of integrability as seen by plotting the variation of the asymptotic gap $\Delta_{a}$ with increasing $\gamma$ for both integrable and non-integrable chaotic systems, considering different values of $h$, as shown in Fig.~\ref{Fig3}(a3-c3). For strong NNN interaction $J_{2}$, it can be observed that for weak $\gamma$, $\Delta_{a}$ is vanishingly small, indicating the presence of a gapless state. Nevertheless, the system gradually acquires a finite gap with increasing $\gamma$ and further increases with it, qualitatively resembling the behavior observed in the integrable model with $J_{2}=0$. However, due to finite-size effects, the gap does not converge precisely to zero, particularly close to the critical boundary $\gamma_c(h) = 4\sqrt{1 - h^2}$ (see Fig.~\ref{Fig3}(c3) for $h = 0.95$), where the region characterized by a gapless spectrum is very small, as can be observed from the phase diagram in Fig.~\ref{Fig1}(a).

These results suggest that even in non-integrable chaotic systems there exists a transition between a gapless and a gapped phase as a function of the measuring rate $\gamma$. This falls in line with the observations of the scaling of the entanglement entropy in the steady state.

\subsection{MODEL \rom{2}}
\label{sec:model2}

In order to further corroborate the results of the previous section, let us now examine the  characteristic features of the steady-state energy spectrum and entanglement of the quantum Ising chain with a uniform longitudinal field, which lacks $\mathbb{Z}_2$ symmetry having in mind the same measurement protocol as outlined in Sec.~\ref{Model_and_Method}. Our aim is to investigate whether the gapless phase, in which the steady-state entanglement spreads logarithmically under weak measurement rates, can emerge in the chaotic region in the absence of a microscopic symmetry in the Hamiltonian; namely, the $\mathbb{Z}_{2}$ symmetry. To address this question, we proceed with measuring the local spins $\hat{\sigma}^{x}_{i}$ in the transverse direction, while considering the following Hamiltonian
\begin{eqnarray}
\hat{\mathcal{H}}_{\rm \rom{2}}=-\sum\limits_{i=1}^{L}\hat{\sigma}^{z}_{i}\hat{\sigma}^{z}_{i+1}-g\sum\limits_{i=1}^{L}\hat{\sigma}^{z}_{i} -h\sum\limits_{i=1}^{L}\hat{\sigma}^{x}_{i},  
\label{QIC_LF}
\end{eqnarray}
where $g$ represents the longitudinal field. 
In the \it no-click limit, \rm the dynamics of the above system under measurements is now governed by the non-Hermitian Hamiltonian, as written below
\begin{eqnarray}
\hat{\mathcal{H}}^{\rm \rom{2}}_{\rm eff}=&-&\sum\limits_{i=1}^{L}\hat{\sigma}^{z}_{i}\hat{\sigma}^{z}_{i+1}-g\sum\limits_{i=1}^{L}\hat{\sigma}^{z}_{i}\notag\\
&-&\big(h+\frac{\imath \gamma}{4}\big)\sum\limits_{i=1}^{L}\hat{\sigma}^{x}_{i},  
\label{H_eff_LF}
\end{eqnarray}

where the imaginary part in the transverse field arises as a reminiscence of the measurement process, similar to previous cases.\\
%%%%%%%%%%%%%%%%Figure 4%%%%%%%%%%%%%%%%%%%%%%
\begin{figure}[H]
\centering
\includegraphics[width=1.02\columnwidth]{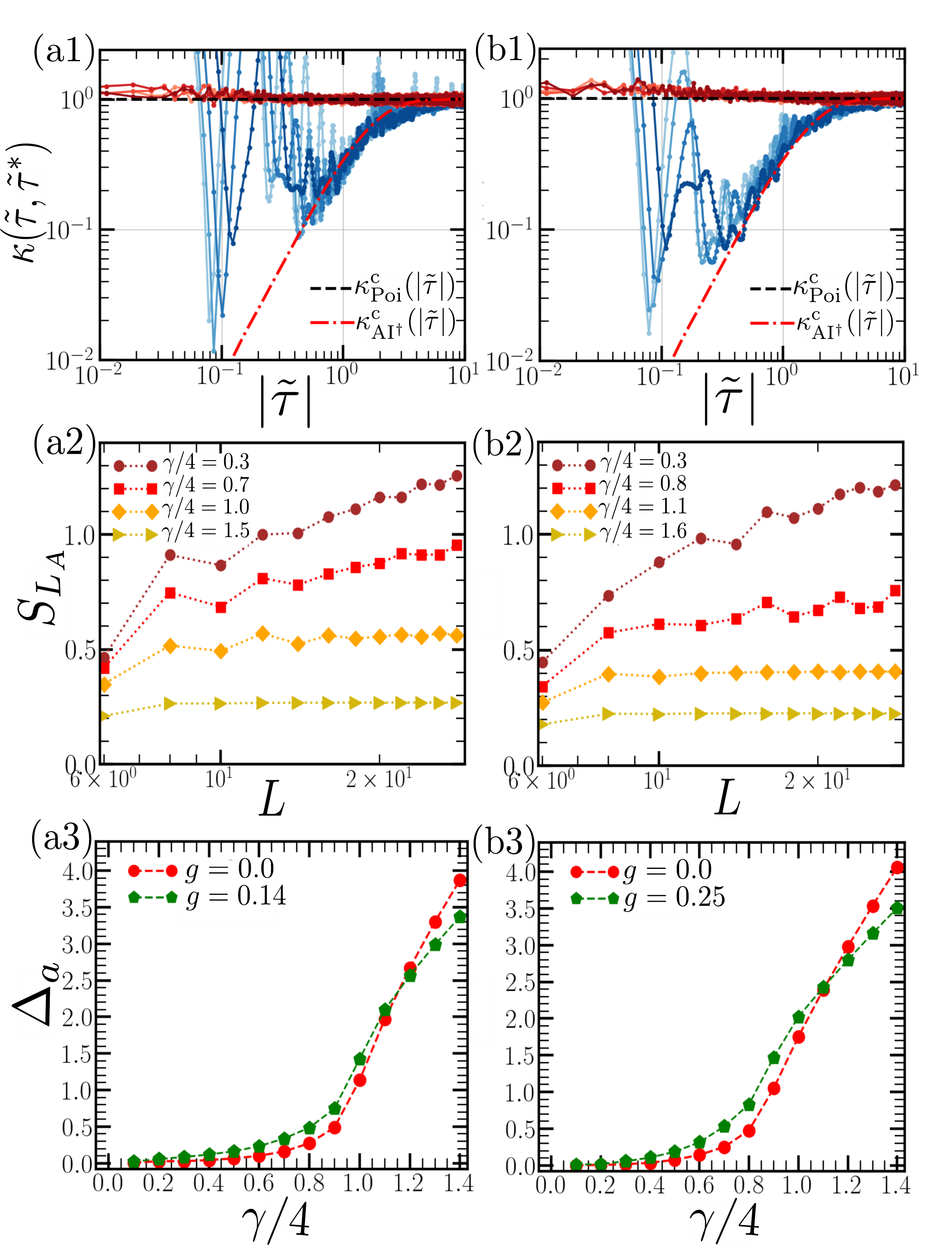}
\caption{The DSFF is depicted as a function of $|\tilde{\tau}|$ for fixed $\theta$ with parameter values: (a1) $h=0.25$, $g=0.14$ (blue shades) and (b1) $h=0.5$, $g=0.25$ (blue shades), while the red shaded lines in both figures represent the DSFF for the integrable model at $g=0$. Different shades of colors correspond to $\theta \in [\pi/20,6\pi/20]$ in steps of $\pi/20$, ordered from lightest to darkest shade. The DSFF is calculated at $\gamma/4=0.3$ by averaging over 1000 realizations of the system with $L=14$. The entanglement entropy $S_{L_A}$ is plotted as a function of $L$ for different measurement rates as indicated in the legend, at the chaotic limit: (a2) $h=0.25, g=0.14$ and (b2) $h=0.5$, $g=0.25$. The variation of the asymptotic gap $\Delta_{a}$ (considering up to $L=26$) is shown as a function of $\gamma$ with transverse field, (a3) $h=0.25$ and (b3) $h=0.5$, considering both integrable (red circles) and chaotic (green pentagons) systems. }
\label{Fig4}
\end{figure}
%%%%%%%%%%%%%%%%%%%%%%%%%%%%%%%%%%%%%%%%%%%%%%
From our analyses, it can be observed that deep within the chaotic limit with sufficiently large $g$, there exists a critical measurement rate $\gamma_{c}$ below which entanglement grows logarithmically with system size. This suggests the emergence of a gapless state in the imaginary part of the complex spectrum. In Fig.~\ref{Fig4}(a1,b1), we calculate the DSFF to check for the chaotic regime in the system and subsequently evaluate the entanglement and the asymptotic gap for the selected parameter values, as illustrated in Fig.~\ref{Fig4}(a2-b3). The observed trends in the entanglement spreading with system size, following a logarithmic law for weak $\gamma$ and stabilizing to a constant value for strong $\gamma$ (see Fig.~\ref{Fig4}(a2,b2)), as well as the variation of $\Delta_{a}$ starting from zero and attaining finite values with higher $\gamma$ (as seen in Fig.~\ref{Fig4}(a3,b3)), provide indications of a transition from a gapless to a gapped state due to the increased measurement rate.\\
Therefore, our findings suggest that non-integrable systems, even those with broken $\mathbb{Z}_2$ symmetry, may display similar qualitative features in steady-state entanglement and the transition between gapless and gapped phases, as observed in integrable models under the same measurement strategy. This observation seems to imply that at least in the \it no-click limit \rm such phase transitions are exclusively driven by the external measurement protocols and are not only independent of microscopic details, as in the previous section, but also of symmetry, conservation laws, and integrability.

In the final section, we extend our analysis to examine the quantum Ising chain under monitoring using an alternative measurement approach.

\section{Measurement of the Longitudinal Magnetization}
\label{Longitudinal}
Let us now turn our attention to a slightly different situation where the \it measurement induced \rm term in the Hamiltonian is the one breaking explicitly its integrability. We can do this by discussing the consequences of altering the measurement basis in the \it no-click \rm dynamics of the Ising chain. Already in the integrable limit the importance of the measurement direction for measurement-induced phase transition is readily outlined. For this sake, consider the effective Hamiltonian Eq.~\eqref{H_eff} with $h=0$:
even in the absence of a transverse field a spectral and associated entanglement transition at a critical measurement rate $\gamma_{c}=4$ is observed~\cite{turkeshi_5,schiro}. If on the contrary, the measurement basis is rotated and the local longitudinal field $\hat{\sigma}^{z}_{i}$ in the upward $z$-direction is measured, the resulting non-Hermitian Hamiltonian in the \it no-click limit \rm takes the form
\begin{eqnarray}
\hat{\mathcal{H}}_{\rm L}=-\sum\limits_{i=1}^{L}\hat{\sigma}^{z}_{i}\hat{\sigma}^{z}_{i+1}-\frac{\imath\gamma}{4} \sum\limits_{i=1}^{L}\hat{\sigma}^{z}_{i}.
\label{Ising_longitudinal}
\end{eqnarray}
In this case, the system is classical and the non-Hermitian vacuum is a product state with all spins aligned in the downward $z$-direction. Therefore the stationary state has zero entanglement for all values of $\gamma$. However, our aim is to examine whether the absence of measurement-induced phase transitions with an altered measurement basis persists in chaotic quantum systems under strong integrability-breaking effects.\\

A simple way to assess how this sensitivity is influenced by breaking integrability is to add a transverse field term $h$ ($h<1$) to Eq.~\eqref{Ising_longitudinal},  driving the system into the quantum chaotic limit
\begin{eqnarray}
\hat{\mathcal{H}}_{\rm L}=-\sum\limits_{i=1}^{L}\hat{\sigma}^{z}_{i}\hat{\sigma}^{z}_{i+1}-\frac{\imath\gamma}{4} \sum\limits_{i=1}^{L}\hat{\sigma}^{z}_{i}-h\sum\limits_{i=1}^{L}\hat{\sigma}^{x}_{i}.
\label{H_eff_long}
\end{eqnarray}
%%%%%%%%%%%%%%%%Figure 5%%%%%%%%%%%%%%%%%%%%%%
\begin{figure}[b]
\centering
\includegraphics[width=0.8\columnwidth]{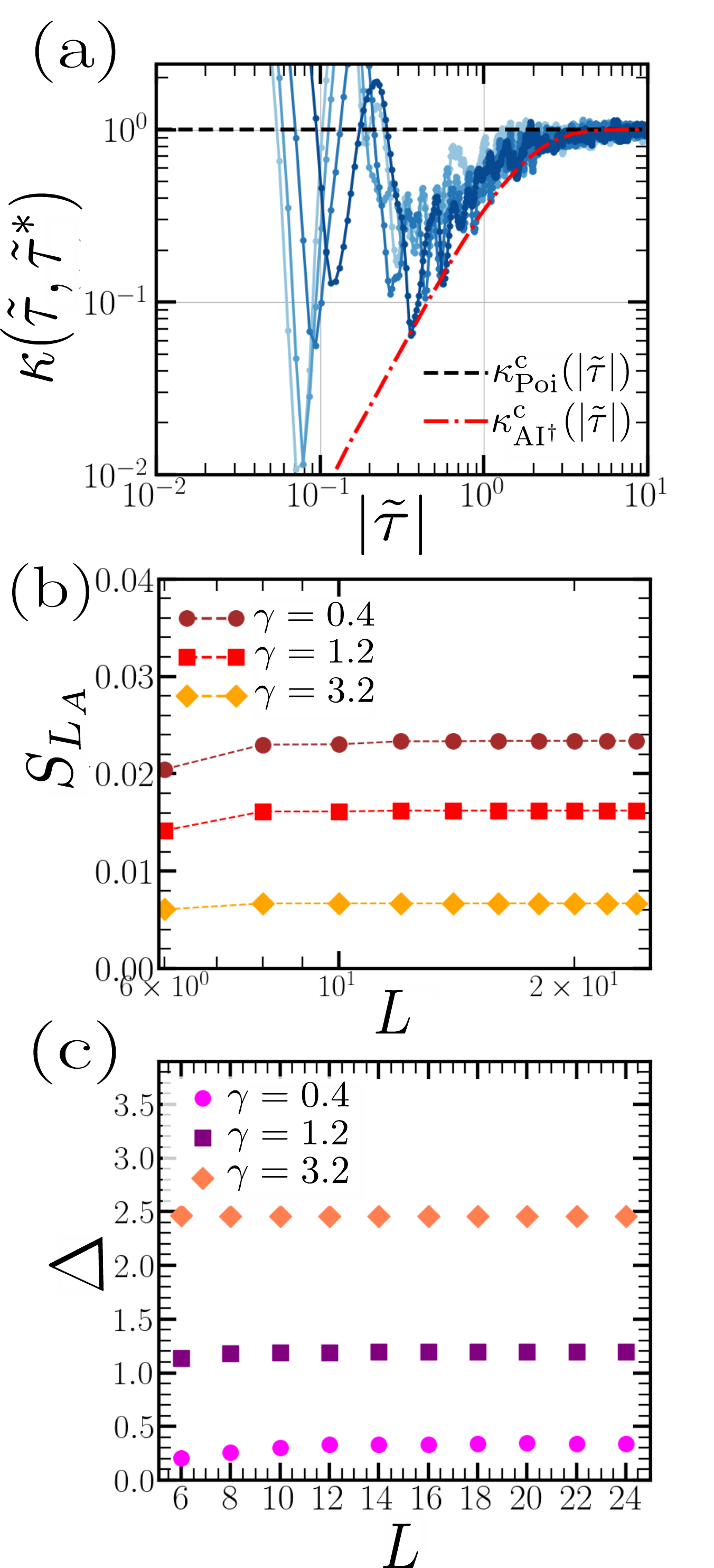}
\caption{(a) The DSFF is shown as a function of $|\tilde{\tau}|$ for fixed $\theta$ at $h=0.7$, $\gamma=1.2$ that agrees with the DSFF $\kappa^{c}_{\rm AI^{\dagger}}(|\tilde{\tau}|)$ calculated for non-Hermitian random matrices under ${\rm AI^{\dagger}}$ symmetry class, denoted by the red dash-dotted line. Different shades of blue correspond to $\theta \in [\pi/20,6\pi/20]$ in steps of $\pi/20$, ordered from lightest to darkest shade. For this computation we consider 1000 realizations of the system with $L=14$. (b) The entanglement entropy $S_{L_A}$ with increasing system size $L$ at $h=0.7$ for different measurement rates. (c) The variation of the spectral gap $\Delta$ with $L$ is illustrated at different values of $\gamma$. }
\label{Fig5}
\end{figure}
%%%%%%%%%%%%%%%%%%%%%%%%%%%%%%%%%%%%%%%%%%%%%%
Together with the lack of integrability this Hamiltonian has the peculiarity of being invariant under $\cal{PT}$ symmetry, meaning that its spectrum is either purely real or, as in the parameter regime considered here, made of complex conjugate pairs of eigenvalues when $\cal{PT}$ symmetry is broken at the level of eigenstates~\cite{olalla09}. It is important to note that, although Eq.~\eqref{H_eff_long} resembles Hamiltonians studied in the context of the Yang-Lee edge singularity, the system remains far from the associated critical point when $h<1$. In contrast, for $h>1$, the system may approach criticality related to the Yang-Lee edge singularity ~\cite{chen}, but this is beyond the scope of our current analysis.

Following the same approach as in the previous sections, we compute the entanglement entropy of the vacuum state $| \emptyset_{\eta} \rangle$ characterized by the largest imaginary part in the complex eigenvalues. We focus on a parameter range where the system is fully chaotic, as ascertained by the DSFF depicted in Fig.~\ref{Fig5}(a). Even in the presence of a transverse field we observe that the entanglement remains bounded with system size, following an area law even for small values of $\gamma$, as seen in Fig.~\ref{Fig5}(b). Such behavior in the entanglement scaling further suggests the absence of any gapless phase for weak $\gamma$, unlike the previous cases. As shown in Fig.~\ref{Fig5}(c), the spectral gap $\Delta$, calculated as the difference between the largest (non-Hermitian vacuum) and the second largest ($1^{\rm st}$ excited state) of the spectrum in the imaginary part ($\Delta=\Gamma^{0l}-\Gamma^{1l}$), no longer vanishes with system size $L$. Instead, it remains constant at some finite value. 

We may further corroborate the numerical results above by considering the perturbative correction to the imaginary part of the spectrum of the non-Hermitian vacuum $| \emptyset_{\eta}\rangle$ resulting from the inclusion of a transverse field $h$ in Eq.~\eqref{Ising_longitudinal}. Initially, for $h=0$, the vacuum state energy, which has the largest imaginary part, is given by $-L+\imath L\gamma^{\prime}\equiv E^{(0)}+\imath \Gamma^{(0)}$ (where $\gamma^{\prime}=\gamma/4$) with all spins aligned in the downward direction. Whereas, the left vacuum state, denoted as $\langle \tilde{\emptyset}_{\eta}|$, is the eigenstate of $H^{\dagger}$ in Eq.~\eqref{Ising_longitudinal} corresponding to energy $-L-\imath L\gamma^{\prime}$ with the smallest imaginary part, satisfying the condition $\langle\tilde{\emptyset}_{\eta}|\emptyset_{\eta}\rangle=1$ \cite{walker}. The first-order correction due to the transverse field is
\begin{equation}
\langle \tilde{\emptyset}_{\eta}| -h\sum^{L}_{i}\hat{\sigma}^{x}_{i} | \emptyset_{\eta}\rangle=0,
\end{equation}
since the operator $\hat{\sigma}^{x}_{i}$ induces localized spin flips at each $i^{\rm th}$ site. The second-order correction can be estimated as:
\begin{align}
&\frac{\sum_{\eta} \langle \tilde{\eta}| -h\sum^{L}_{i}\hat{\sigma}^{x}_{i} | \emptyset_{\eta}\rangle \langle \tilde{\emptyset}_{\eta}| -h\sum^{L}_{i}\hat{\sigma}^{x}_{i} |\eta \rangle}{(E_{\emptyset_{\eta}} - E_{\eta})} \notag \\
&= \frac{-Lh^{2}(2 + \imath \gamma^{\prime})}{(8 + 2\gamma^{\prime 2})} \equiv E^{(2)} + \imath \Gamma^{(2)},
\end{align}.
where $|\eta \rangle$ and $|\tilde{\eta}\rangle$ denote the right and left excited states, respectively, with one spin flipped at the $i^{\rm th}$ site. Considering $h<1$, it becomes evident that $\Gamma^{(2)}<<\Gamma^{(0)}$, indicating  that the imaginary part of the spectrum always remains gapped despite the inclusion of $h$. Therefore, the presence of a gapped phase throughout the entire parameter range directly suggests an area law in entanglement scaling. This is because the localized perturbation resulting from $h$ introduces a finite correlation length $\xi$ into the system, leading to an exponential decay in the correlation function. As a result, the probability of finding a spin flip at a distance $|i-j|$ from another spin decreases exponentially, keeping the entanglement bounded. This phenomenon observed in the non-Hermitian vacuum is therefore analogous to the area law conjecture \cite{hastings, plenio_area} of ground state entanglement in Hermitian quantum Ising chain away from criticality. 

In conclusion the arguments above demonstrate that while the emergence of spectral and entanglement transitions in the \it no-click limit \rm of a monitored quantum Ising chain appears to depend crucially on the choice of measurement basis, a detailed analysis of this issue, both numerical and using previous analytical results on Eq.~\eqref{H_eff_long} will be the subject of further studies.

\section{Conclusions}
\label{Conclusion}
In this work, we have performed exact numerical simulations to study the robustness of measurement-induced phase transitions in the \it no-click limit \rm of the quantum Ising chain under various non-integrable perturbations, applying different measurement protocols. Using exact diagonalization, by which a clear transition between logarithmic to constant scaling of the entanglement entropy in the integrable chain is seen, we investigate the effects of breaking integrability by either a  next-nearest neighbor ferromagnetic interaction or a longitudinal field, breaking the $\mathbb{Z}_{2}$ symmetry,  by first analyzing the Dissipative Spectral Form Factor (DSFF) as a measure of chaos in non-Hermitian systems. Interestingly, we found that the behavior of the spectral gap, as well as the associated transition in steady-state entanglement scaling with the measurement rate in the chaotic system, remains qualitatively similar to that observed in the integrable model.

These results suggest that the emergence of the subextensive critical phase, characterized by a gapless imaginary part of the spectrum in the free fermionic Hamiltonian under a weak measurement rate, remains robust irrespective of both integrability and the preservation of $\mathbb{Z}_2$ symmetry. On the other hand, we observed that this gapless critical phase disappears, and the entanglement conforms to an area-law scaling regardless of the measurement rate when the measurement basis is altered. Thus, the interactions and symmetry of the system do not fundamentally alter the qualitative nature of the measurement-induced phase transition; rather, the direction of measurement proves crucial in determining the onset of these transitions. These findings further imply that the entanglement characteristics of the non-Hermitian vacuum can be regarded as an extension of the ground state entanglement observed in Hermitian systems, as discussed in Ref.~\cite{silva_2,turkeshi_2}. Lastly, in relation to this study, it would be intriguing to investigate how interactions, symmetry, and the choice of measurement basis affect the entanglement and spectral features of the monitored quantum Ising chain under generic trajectories involving quantum jumps, which remains a subject for further investigation. Moreover, our framework can be extended to study higher-dimensional systems, different symmetry sectors, long-range interactions, and the presence of disorder, further generalizing the
study of MIPT in chaotic regimes.

\begin{acknowledgments}
M.M. thanks A. Paviglianiti for discussions. D.S. acknowledges support from an NSERC Discovery grant and the Canada Research Chair Program. The work of M.B. was partially supported by an Alliance International Catalyst Quantum grant. A.S. and M.M. acknowledge the support of the grant PNRR MUR project PE0000023-NQSTI. A.S. acknowledges support of the project “Superconducting quantum-classical linked computing systems (SuperLink)”, in the frame of QuantERA2 ERANET COFUND in Quantum Technologies.
Computations were performed on the Ulysses v2 high performance computer at SISSA and the Niagara supercomputer at the SciNet HPC Consortium. SciNet is funded by: the Canada Foundation for Innovation; the Government of Ontario; the Ontario Research Fund - Research Excellence; and the University of Toronto. 
\end{acknowledgments}

\section*{Data Availibility Statement}The data that support the findings of this study are available from the authors upon reasonable request.

\bibliography{refs}

\end{document}